\documentclass[structabstract]{aa}
\usepackage{graphicx}
\def\spose#1{\hbox to 0pt{#1\hss}} 
\def\simlt{\mathrel{\spose{\lower 3pt\hbox{$\mathchar"218$}} 
\raise 2.0pt\hbox{$\mathchar"13C$}}} 
\def\simgt{\mathrel{\spose{\lower 3pt\hbox{$\mathchar"218$}} 
\raise 2.0pt\hbox{$\mathchar"13E$}}}

\begin{document}
\title{Galactic Constraints on Supernova Progenitor Models}
\author{I.~A. Acharova, 
\inst{1}\,\
B.~K. Gibson,
\inst{2}\,\
Yu.~N. Mishurov
\inst{1}
and
V.~V. Kovtyukh\inst{3}}
%
%
\institute{Department of Physics, Southern Federal University, 
5 Zorge, Rostov-on-Don, 344090, Russia\\
\email {iaacharova@sfedu.ru (IAA); unmishurov@sfedu.ru (YuNM)}
\and 
Jeremiah Horrocks Institute, University of Central Lancashire, Preston, 
PR1~2HE, UK\\
\email{brad.k.gibson@gmail.com}
\and 
Astronomical Observatory, Odessa National University, and Isaac Newton 
Institute of Chile, Odessa Branch, T.G. Shevchenko Park, 65014, 
Odessa, Ukraine\\
\email {val@deneb1.odessa.ua}
}
\date{Received \today; accepted }
\authorrunning{Acharova et~al.}
\titlerunning{Galactic Constraints on Supernova Models}

\abstract
{}
{To estimate the mean masses of oxygen and iron ejected per each type of 
supernovae (SNe) event from observations of the elemental abundance patterns
in the Galactic disk and constrain the relevant SNe progenitor models.}
{We undertake a statistical analysis of the radial abundance 
distributions in the Galactic disk within a theoretical framework for 
Galactic chemical evolution which incorporates the influence of spiral 
arms. Such a framework has been shown to recover the non-linear 
behaviour in radial gradients, and enables one to estimate the mean 
masses of oxygen and iron ejected during SNe explosions, and place 
constraints on SNe progenitor models.}
{(i) The mean mass of oxygen ejected per core-collapse SNe (CC~SNe) event 
(which are concentrated within spiral arms) is $\sim$0.27~M$_{\odot}$; 
(ii) the mean mass of iron ejected by `tardy' Type~Ia SNe (SNeIa; progenitors 
of whom are older/longer-lived stars with ages $\simgt$100~Myr and up to several 
Gyr, which do not concentrate within spiral arms) is $\sim$0.58~M$_{\odot}$; 
(iii) the upper mass of iron ejected by prompt SNeIa (SNe whose progenitors 
are younger/shorter-lived stars with ages $\simlt$100~Myr, which are concentrated 
within spiral arms) is $\leq$0.23~M$_{\odot}$ per event; 
(iv) the corresponding mean mass of iron produced by CC~SNe is 
$\leq$0.04~M$_{\odot}$ per event; 
(v) short-lived SNe (core-collapse or prompt SNeIa) supply $\sim$85\% of the 
Galactic disk's iron.}
{The inferred low mean mass of oxygen ejected per CC~SNe event implies a
low upper mass limit for the corresponding progenitors of $\sim$23~M$_{\odot}$, 
otherwise the Galactic disk would be overabundant in oxygen. This inference
is the consequence of the non-linear dependence between the upper limit of the 
progenitor initial mass and the mean  mass of oxygen ejected per CC~SNe 
explosion. 
The low mean mass of iron ejected by prompt SNeIa, relative 
to the mass produced by tardy SNeIa ($\sim$2.5 times lower), prejudices 
the idea that both sub-populations of SNeIa have the same physical 
nature. We suggest that, perhaps, prompt SNeIa are more akin to CC~SNe, 
and discuss the implications of such a suggestion.}
\keywords{Nuclear reactions. nucleosynthesis, abundances -- stars: supernovae
-- Galaxy: evolution
}

\maketitle

\section{Introduction} 
Supernovae (SNe) play a pivotal role in driving our understanding of the 
cosmology of the Universe and the formation of life within it. SNe supply 
the surrounding interstellar medium (ISM) with energy, cosmic rays, and 
the heavy elements necessary for building planets and complex biological 
life (Lineweaver et~al. 2004). The knowledge that a sub-set of Type~Ia SNe 
(SNeIa) could be considered as standard candles, allowed them to be used 
as `standard candles', resulting in the determination of both the 
Universe's present-day expansion rate (Gibson et~al. 2000) and its 
acceleration (Riess et~al. 1998; Perlmutter et~al. 1999), thereby 
providing evidence for the existence of `dark energy'.

While tempting to think of SNeIa as a uniform population, with 
progenitors of ages typically on the order of several billion years, 
more recent works suggest that they consist of two sub-populations:
(i) `prompt' SNeIa (denoted hereafter as SNeIa-P) -- their progenitors 
are relatively short-lived stars of ages of several tens of millions of 
years up to $\sim$100~Myr, and (ii) `tardy' SNeIa (denoted hereafter as 
SNeIa-T) -- their progenitors are relatively longer-lived systems with 
ages from $\sim$100~Myr up to several Gyrs (Matteucci \& Greggio 
1986; Bartunov et~al. 1994; Mannucci et~al. 2005, 2006; Maoz et~al. 
2010; Li et~al. 2011). This `diversity' means cosmologists have had to 
be careful in their application of SNeIa as standard candles. Indeed, 
SNeIa-P are brighter objects than tardy ones (Mannucci et~al. 2005; 
Sullivan et~al. 2006), and have therefore been more readily seen in 
distant galaxies.  If the prompt SNeIa did not obey the `pattern' 
usually associated with SNeIa, they may have distorted the observed 
inferences, due to selection effects.

Core-collapse SNe (CC~SNe) progenitors are known to be of higher mass than 
those of SNeIa, but whether they are associated with the \it most \rm 
massive stars has been called into question, with analyses of their 
pre-SNe progenitors suggesting masses less than about 20~M$_\odot$ (e.g., 
Kochanek et~al. 2008; Smartt et~al. 2009).  The latest developments in the 
field have unfortunately not clarified the situation. For example, Brown 
\& Woosley (2013) suggest that stars of masses up to 120 M$_{\odot}$ 
should explode as CC~SNe, based upon the assumption that the $^{16}$O 
abundance of the Sun corresponds to the typical value encountered in the 
Galactic disk. Conversely, Eldridge et~al. (2013) argue persuasively for a 
much lower CC~SNe upper mass limit.

It may seem surprising at first, but chemical fingerprints held within the 
Galaxy's disk may impose important, unforeseen, constraints on the breadth 
of SNe progenitors.  We will show that the mean masses of oxygen and iron 
ejected per CC~SNe event and iron per SNeIa-P appear to be significantly 
lower than usually assumed. A lower mean ejected mass of oxygen per CC 
event supports the inferences of Heger et~al. (2003), Kochanek et~al. 
(2008), Smartt et~al. (2009), Moriya et~al. (2011), and others, who 
suggest that extremely massive stars do not explode as CC~SNe (and 
therefore do not pollute the surrounding ISM with heavy elements), in 
order to avoid an overproduction `problem' for oxygen within the Galaxy. 
However, if that is the case, we are faced with a conundrum, in that 
present observations certainly demonstrate the existence of very massive 
stars, up to $\sim$~100~M$_{\odot}$ (e.g., Schnurr et~al. 2008; de Mink 
et~al. 2009; Crowther 2010). Can we state with certainty that these very 
massive stars will not end their lives as CC~SNe? In this sense, and being 
counter to conventional wisdom governing CC~SNe searches, a strategy for 
searching events associated with a massive star disappearing `quietly', 
rather than in a spectacular explosion, appears an interesting approach 
(Kochanek et~al. 2008).

Canon suggests that sub-luminous SNeIa produce little in the way of nickel 
(and hence iron, via radioactive decay), while luminous SNeIa eject 
significantly more (e.g. Gonzalez-Gaitan et~al. 2011; Truran et~al. 2012). 
However, as we will demonstrate, our results suggest that the mean mass of 
iron ejected per SNeIa-P is lower than that produced per SNeIa-T, by a 
factor of $\sim$2.5. The low ejected masses of iron from SNeIa-P, along 
with the extremely low one produced by CC SNe, encourages the suggestion 
that the nature of prompt SNeIa is, likely, dissimilar to that of the 
standard model of SNeIa.

\section{Methodology}
The basis of Galactic chemical evolution is predicated upon theoretical 
studies of pre-SN stellar evolution, and in particular the predicted yield 
of a given isotope, much as we have adopted in our earlier work (Gibson 
1997; Mishurov et~al. 2002; Acharova et~al. 2005ab,2010,2011,2012; Lewis 
et~al. 2013).

In the present paper, we re-formulate this classical approach and instead 
\it derive \rm the mean ejected masses of oxygen and iron making using of 
an extensive observational dataset of Cepheid abundances (Acharova et~al. 
2012, hereafter AMK), the frequencies of various SNe sub-type event rates 
from the LOSS survey (Li et~al. 2011), and refined statistical methods for 
the analysis of the non-linear radial distributions of oxygen and iron in 
the Galactic disk, as per Acharova et~al. (2011, hereafter AMR) and AMK.

Oxygen was chosen for this work as its radial distribution demonstrates a 
distinct `feature' in its distribution -- a sharp `bend' in the slope of 
the distribution in moving from the inner (relative to the Sun) part of 
the disk to the outer part (see below). Besides, oxygen is mainly produced 
by CC~SNe which are concentrated within spiral arms; as shown in earlier 
papers in this series, the so-called co-rotation resonance of Galactic 
spiral density waves with rotating matter of the disk is responsible for 
this feature. Hence, oxygen can be considered as something of a `clean' 
indicator of the spiral arms' influence on the radial distribution of 
heavy elements in the Galactic disk.

Conventional wisdom suggests that $\sim$60$-$70\% of iron in the Galaxy 
was produced by SNeIa (Gibson 1998, and references therein).  If all SNeIa 
were associated with old stars alone, we would not expect to see any 
obvious feature in the radial distribution of iron.  This picture changed 
with the work of Andrievsky et~al. (2002abc) who showed that iron also 
demonstrates an inflection, albeit not as sharp as that seen for oxygen, 
but the change in its radial gradient is noticeable and located close to 
the bend seen in the oxygen distribution. Such a coincidence was difficult 
to explain, initially, since old stars do not concentrate in spiral arms 
(Mishurov \& Acharova 2011), but the discovery of two sub-populations of 
SNeIa (as noted in \S1) provided a means by which to explain this apparent 
problem.  Such `features' enable us to decompose the contributions of 
various sources of iron synthesis.

Finally, having the mean ejected masses of oxygen, we derive constraints 
on the upper masses of stars exploding as CC~SNe using a technique similar 
to that used by Gibson (1998). Contrary to the aforementioned conventional 
wisdom, we now infer that the short-lived SNe (i.e., CC~SNe or SNeIa-P) 
supply $\sim$85\% of the Galaxy's iron (cf. $\sim$30$-$40\%, from our 
earlier work: Gibson 1998; Matteucci 2004; Acharova et al. 2010; AMK).

\subsection {{\it Observational Data}}
Fig.~1 illustrates the aforementioned feature, as seen in the Milky Way's 
radial ($r$) distributions of oxygen\footnote{ Similar `structure' in 
the radial oxygen abundance gradient is seen in both M83 (Bresolin et~al. 
2009) and NGC~5668 (Marino et~al. 2012).} and iron. Here, 
$[X/H]=log(N_X/N_H)_{star}-log(N_X/N_H)_{\odot}$, where $N_{X,H}$ is the 
number of atoms of element $X$ or hydrogen, respectively. The angle 
brackets ($y$-axis), $<...>$, mean that we have divided $r$ into bins of 
500~pc width and averaged the corresponding data within each bin.

The observational material is based on spectroscopic data for 283 
classical Cepheids (872 spectra in total; abundances given by AMK in their 
Table~2). Cepheids have been employed due to their intrinsic luminosities 
being sufficient to allow their observation at significant distances from 
the Sun; they also possess precise distances, and are sufficiently young 
as to represent abundances in the ISM at the time and location of their 
birth.  For the given sample, both iron and oxygen abundance 
determinations exist for the bulk of the Cepheids.

\begin{figure}
\includegraphics {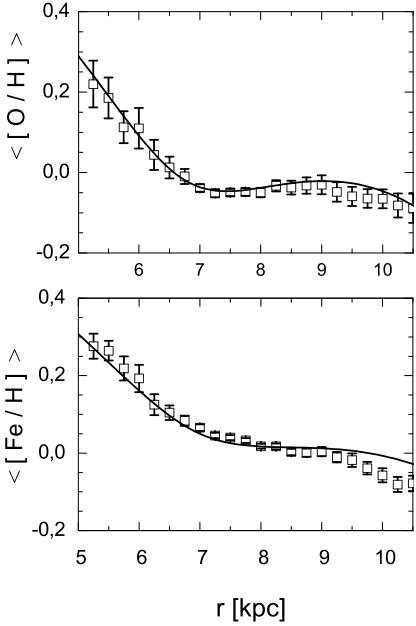}
\caption{Open squares are the averaged observed radial distributions of 
oxygen (upper panel) and iron (bottom panel) derived using Cepheids in 
the Milky Way (data taken from AMK). The error-like bars are the 
standard deviations of the corresponding mean values. The solid lines 
are the best fitted theoretical radial abundance distributions (see \S3.1).}
\label{f1}
\end{figure}

The radial distribution of oxygen demonstrates a sufficiently sharp break 
in its behaviour near $r$$\sim$7~kpc (for the solar Galactocentric 
distance, we adopt $r_0$$\equiv$7.9~kpc). However, for iron, there is no 
such sharp `bend' in the gradient at the same distance, although it is 
still clear that its distribution cannot be satisfactorily described by a 
singular linear function. In what follows, we restrict ourselves to radii 
$r$$\leq$10.5~kpc.

\subsection {{\it Equations for the Chemical Evolution of the ISM}} 
The formalism employed to model the chemical evolution of the Galactic 
disk is based upon the classical 1D (radial) approach of Tinsley (1980):

\begin{equation}
\dot{\mu}_g=f-\psi + \int\limits_{m_L}^{m_U}{(m-m_w)\psi(t-\tau_m)\phi (m)\,dm},
\end{equation}
\noindent

\begin{eqnarray}
{\dot{\mu}_j}&=&\int\limits_{m_L}^{m_U}{(m-m_w)\,Z_j(t-\tau_m)
\psi(t-\tau_m)\phi 
(m)\,dm} + \nonumber\\ &&E_j^{\rm Ia}+E_j^{\rm cc} + fZ_{j,f}-Z_j\psi
+\frac 
{1}{r}\frac {\partial {}}{\partial r}\left (r\mu_g D\frac {\partial 
Z_j}{\partial r}\right).
\end{eqnarray}  
\noindent
Here $\mu_g$ is the surface mass density of the interstellar gas (ISG), 
$f=f_0\exp(-r/r_d-t/t_f)$ is the infall rate of intergalactic gas onto the 
Galactic disk with temporal and radial scales of $t_f=2$~Gyr (AMR) and 
$r_d=3.5$~kpc (Marcon-Uchida et~al. 2010), respectively. $f_0$ is computed 
using the normalizing condition that the total surface density of the 
Galactic disk (stellar + gaseous) at the Sun's galactocentric distance 
$r_0$ at the present epoch ($t=T_D=10$~Gyr, where $T_D$ is the age of the 
disk), is 50~M$_{\odot}$~pc$^{-2}$ (Haywood et~al. 1997; Portinari \& 
Chiosi 1999). $\psi$ is the star formation rate (SFR), $m$ the stellar 
mass (in solar mass units), $\tau_m$ the lifetime of a star of mass $m$ 
(in Gyr), where $\log(\tau_m)=0.9-3.8\log(m)+\log^{2}(m)$ (Tutukov \& 
Kruegel 1980; cf. Fig~1 of Gibson 1997), and $\phi(m)$ the adopted Kroupa 
et~al. (1993) initial mass function (IMF). $m_w$ is the mass of stellar 
remnants (white dwarfs, neutron stars, and black holes): for $m \le 10$, 
$m_w = 0.65m^{1/3}$; in the range $10 < m < 30$, $m_w = 1.4$ (neutron 
star); for $30 \le m < m_U$, the remnant is assumed to be a black hole of 
$m_w = 10$; finally, for $m \ge m_U=70$ the stars are assumed to collapse 
to black holes immediately after their birth and removed from future 
chemical evolution (Tsujimoto et~al. 1995).  The lower stellar mass limit 
is taken to be $m_L = 0.1$~M$_\odot$, $\mu_j$ is the surface mass density 
of oxygen or iron (`{\it j=O}' or `{\it j=Fe}', respectively), 
$Z_j=\mu_j/\mu_g$ is the mass fraction for the elements in the 
interstellar medium, $Z_{j,f}$ is the metallicity of the infalling gas, 
while $t$ is time. The last term in equation (2) describes the radial 
diffusion of the elements, the expression for the diffusion coefficient, 
$D$, being given by Acharova et~al. (2010).

We take the `edge' of the Galactic disk to be $r =R_G \equiv 25$~kpc; for 
the initial conditions, we assume $\mu_g(t=0)=0$ and $Z_j(t=0)=Z_{j,f}$. 
We experimented with various initial fractions ($Z_{O,f}/Z_{O,\odot}$ = 
0.02, 0.05, and 0.1, with the adopted fraction for iron being 
$Z_{Fe,f}=0.02Z_{Fe,\odot}$. Parameter space studies such as these 
allow us to assess the sensitivity of the results - especially the 
predicted [O/Fe] ratio - to the conditions at the epoch of Galactic disk 
formation (e.g., Renda et~al. 2005; Fuhrmann \& Bernkopf 2008).

The enrichment rates, $E^{\rm i}_j$, of the Galaxy for the $j$th chemical 
element due to the explosion of the $i$th type of SNe~ -- CC SNe (`i = 
cc') or SNeIa (`i = Ia') are given by the expression (Tinsley 1980):
\begin{equation}
E_j^{\rm i}(r,t)= P_j^{\rm i} R^{\rm i}(r,t),
\end {equation}
\noindent
where $R^{\rm i}$ are the rates of the $i$th type of SNe explosions per 
unit of surface area, and $P^{\rm i}_j$ are the mean masses of the $j$th 
element ejected per $i$th type of SN event. The above mean ejected masses, 
$P_j^{\rm i}$, are our target free parameters.

To close the equations, a Schmidt--like approximation for the star 
formation is often adopted (i.e., $\psi \propto \mu_g^k$, with 
$k$$\sim$1.5 : Kennicutt 1998). However, to explain the formation of the 
`structure' in the radial gradient of oxygen (and iron), we adopt a 
formalism for $\psi$ which explicitly takes into account the effects of 
the spiral arms.  We follow the prescription proposed by Wyse \& Silk 
(1989, hereafter WS) and Portinari \& Chiosi (1999, hereafter PC) which is 
based on the popular suggestion that Galactic spiral shocks stimulate star 
formation (e.g., Roberts 1969; Shu et~al. 1972). In this prescription, the 
functional form follows $\psi \propto |\Omega(r) - \Omega_P|\mu_g^k$, 
where $\Omega(r)$ is the angular rotation velocity of the Galactic matter 
and $\Omega_P$ is the rotation velocity of the Galactic spiral waves 
responsible for the arms. WS and PC assume that: (i) Galactic spiral 
shocks stimulate formation of stars of all masses (both high and low), and 
(ii) the co-rotation resonance, $r_c$, where $\Omega(r_c)=\Omega_P$, is 
situated at the edge of the Galactic disk (Lin et~al. 1969). In what 
follows, we adopt the aforementioned equations of galactic chemical 
evolution, including the effects of spiral arms, under the assumption that 
spiral arms stimulate formation of sufficiently massive stars, but now 
considering the case for which the corotation resonance is situated close 
to the Sun.\footnote{Such a model for the Galaxy's spiral density wave 
pattern was first proposed by Marochnik et~al. (1972) and Cr\'ez\'e \& 
Mennesier (1973); see also Mishurov et~al. (1979,1997,1999) and L\'epine 
et~al. (2001).}

\subsection{Formation of the Galactic Gaseous Disk}
Until recently, little consensus existed as to the role of spiral arms 
in triggering star formation across the full spectrum of stellar masses. 
For our work, though, what is important is whether or not the 
sources of heavy elements are concentrated within the arms at the moment 
of ejection of the synthesized elements to the surrounding ISM.
Observations do demonstrate that CC~SNe \it are \rm strongly associated 
with the spiral arms (Bartunov et~al. 1994; Li et~al. 2011). Progenitors 
of CC~SNe are massive stars ($m>8$~M$_\odot$) with very short lifetimes 
($\tau$$\simlt$20~Myr). As such, they have not had sufficient time to 
move significantly from their birth location within the arm. Conversely, 
as noted in \S1, the progenitors of SNeIa-T have lifetimes in excess of 
$\sim$100~Myr (up to several Gyr or more - Matteucci \& Greggio 
1986).  Even if these (lower mass) progenitors had been born in the 
spiral arms, by the time of their explosion, they now appear uniformly 
distributed in Galactic azimuth (Mishurov \& Acharova 2011). Finally, 
SNeIa-P are strongly associated with the spiral arms (Bartunov et~al. 
1994; Li et~al. 2011) and the ages of their progenitors may be 
estimated to be $\simlt$100~Myr.\footnote{To cross an 
interarm distance, it takes the time interval $\sim~\pi |\Omega - 
\Omega_P|^{-1} \sim 100$ Myr. Hence we can adopt the above value as the 
boundary one separating the prompt short-lived sub-population of SNeIa 
which concentrates in spiral arms from tardy long-lived SNeIa which do 
not concentrate in arms.} (Mannucci et~al. 2005,2006; Matteucci 
et~al. 2006; Maoz et~al. 2010; cf. Matteucci \& Greggio 1986).

Since the mass of a single star with lifetime $\sim$100~Myr is 
$\sim$4~M$_{\odot}$, we split the star formation rate (SFR) into the rates 
of high, $\psi_H$, and low, $\psi_L$, stellar mass formation for 
$m>4$~M$_\odot$ and $m<4$~M$_\odot$, respectively.  Using $\psi\equiv 
\psi\int_{m_L}^{100}m\phi(m)~dm$, we re-cast the second term in the right 
hand-side of equation (1) as:
\begin{equation}
\psi = \psi_{L}\int _{m_L}^{4}m\phi dm + \psi_{H}\int_{4}^{100}m\phi dm.
\end{equation}
\noindent
Here, $m = 100$~M$_\odot$ is the upper mass limit for stars at the time of 
their birth (Romano et~al. 2005), whereas $m_U = 70$ in equations (1) and 
(2) is the upper mass of stars which take part in the Galactic matter 
circulation.

In the classical 1D approach to chemical evolution, it is impossible to 
consider the scattering of long-lived stars over the Galactic disk (cf. 
Mishurov \& Acharova (2011). As such, we assume that low mass stars are 
not concentrated within spiral arms (a fairly conservative assumption 
given the reaosnably long lifetimes of these stars). What this means is 
that for $\psi_{L}$ we use the usual star formation formalism:

\begin{equation}
\psi_{L}=\nu\mu_g^k(r,t),
\end{equation}
where $\nu$ is a normalizing coefficient (see below).

As was shown by Bartunov et~al. (1994) and Anderson et~al. (2012), CC~SNe 
are tightly linked with HII regions and their azimuthal distributions are 
consistent with density and/or galactic shock waves (e.g., Roberts 1969; 
Boeshaar \& Hodge 1977). Hence, the formalism for $\psi_{H}$ may be 
written as $\psi_{H}= \Pi \mu_g^k$, where $\Pi = \alpha\exp [-(\theta - 
\theta_s)/\delta]$ represents the profile of the galactic shock with 
azimuth angle $\theta$, $\alpha$ is the peak value of $\Pi$ at the shock 
front, $\theta_s$ is the location of the shock, and $\delta$ is the 
typical width of an arm in units of Galactic azimuth (Fig~2). Such a 
formalism is valid for $\theta_s < \theta < \theta_s+\pi$; for other 
angles, the expression for $\Pi$ is derived by means of a periodicity 
condition.

\begin{figure}
\includegraphics {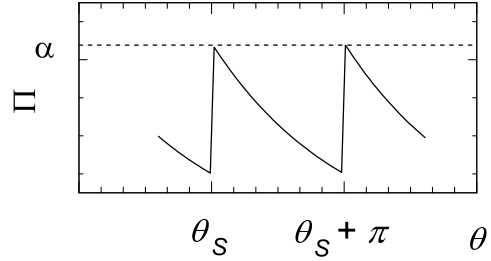}
\caption{The schematic dependence of $\Pi$ on $\theta$, where $\theta_s$ 
is the location of the shock front and $\alpha$ is the peak value of 
$\Pi$ at the shock front.}
\label{f2}
\end{figure}

To pass from 2D ($r,\theta$-representation) to 1D ($r$-representation), 
we average $\psi_{H}$ over $\theta$ and derive: $$<\Pi> = 
\frac{1}{2\pi}\int\limits_{\theta}^{\theta + 2\pi}\Pi d\theta = \alpha 
\frac{\delta}{\pi}(1-e^{-\pi/\delta}).$$ The shock intensity is 
well-approximated by $\alpha \propto r |(\Omega - \Omega_P)|$ (Roberts 
et~al. 1975) and $\delta$ can be estimated as $\delta \approx 
(d/r)\sin(p)$, where $p$ is the pitch angle of a spiral arm and $d$ is 
the typical width of an arm in the direction perpendicular to it. It is 
easy to show that $\delta < 1$. This then yields the same approximation 
for $\psi_{H}$ which was proposed by WS and PC:
\begin{equation}
\psi_{H}=\beta|\Omega(r)-\Omega_P|\mu_g^k.
\end{equation}
Here, the coefficient $\beta$ includes the constants of proportionality 
which enter the above model representations.

In the end, equation (1) can be reduced to the following form:
\begin{eqnarray}
\dot{\mu}_g(r,t)&=&f(r,t)-\psi_{L}(r,t)\int _{m_L}^{4}m\phi dm + \nonumber\\
&&+\int\limits_{m_L}^{m_U}{(m-m_w)\psi_{L}(r,t-\tau_m)\phi (m)\,dm}- \\
&&-\psi_{H}(r,t)[\int\limits_{m_U}^{100}\phi(m) dm + \int\limits_{4}^{m_U}m_w\phi(m) dm]\nonumber.
\end{eqnarray}
\noindent
Here, for massive stars we use explicitly the instantaneous recycling 
approximation.

In the computations which follow, we use the rotation curve based upon 
that of Clemens (1985), adjusted for the adopted scale:
$$r\Omega(r) = 260\cdot \exp\{-[\frac{r}{150}+(\frac{3.6}{r})^2]\} +  
              360\cdot \exp[-(\frac{r}{3.3}+\frac{0.1}{r})].$$
The rotation curve with the adopted $\Omega_P=33$ km s$^{-1}$ kpc$^{-1}$ 
is shown in Fig~3. The corotation resonance is situated at $r_c \sim 
7$~kpc (AMR, AMK). In the expression for $\psi_{H}$, we introduce a 
cut-off factor at the outer Lindblad resonance, $r_{out}$, which 
exponentially decreases beyond the resonance with a scale $\sim$0.5~kpc, 
whereas following WS and PC we do not restrict the region of massive 
star formation by the inner Lindblad resonance.\footnote{Khoperskov 
et~al. (2012) showed that the inner Lindblad resonance has little 
influence on spiral density wave generation.}

\begin{figure}
\includegraphics {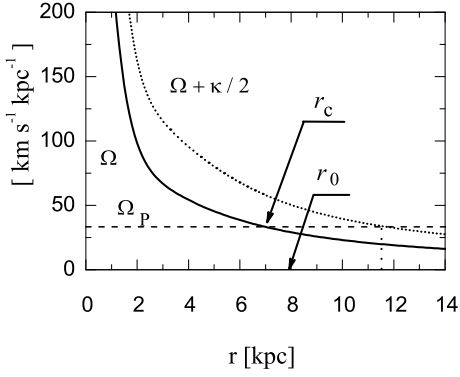}
\caption{The rotation curve (solid line) plotted alongside 
$\Omega+\kappa/2$ (dotted line, where $\kappa$ is the epicyclic 
frequency). The vertical dotted line shows the location of the outer 
Lindblad resonance (at $r_{out}\sim$11.5~kpc).}
\label{f3}
\end{figure}

Equations (5) to (7) have two free parameters: $\nu$ and $\beta$. They 
are fitted so as to ensure the solution satisfies the normalizing 
conditions: (i) $\mu_g(r_0,T_D) = \mu_{g0}=10$ M$_{\odot}$ pc$^{-2}$ 
(Haywood et~al. 1997), and (ii) the computed theoretical frequency for 
CC~SNe events in our Galaxy at the present epoch, $F^{\rm cc}_{th}$, 
equates to the observed value: $F^{\rm cc}_{obs}\sim 2.3$ per century 
(Li et~al. 2011).

The theoretical frequencies for the $i$th type of SNe event, $F^{\rm 
i}_{th}$, are computed through the corresponding rates $R^{\rm i}$ as:
\begin{equation}
F^{\rm i}_{th} = 2\pi \int\limits_{0}^{R_G} R^{\rm i}(r,T_D)\,r dr,
\end{equation}
\noindent
where $R_G\equiv$25~kpc is the adopted radial extend of the Galaxy
and the rate for CC~SNe events is:
\begin{equation}
R^{\rm cc}(r,t)= \psi_{H}(r,t) \int\limits_{8}^{m^{cc}_U}\phi (m)\,dm,
\end{equation}
where $m^{cc}_U$ is the upper limit to the initial mass of 
CC~SNe progenitors. Since its value is not known {\it a priori}, 
initially we adopt $m^{cc}_U = m_U$.

Finally, to solve the above equations and derive $\mu_g(r,t)$ we employ 
the following iterative procedure. At the first step, we suppose that 
$\beta = 0$ and solve the corresponding equations for a set of $\nu$, 
seeking the minimum of the discrepancy 
$\Delta_\mu=[\mu_g(r_0,T_D)-\mu_{g0}]^2$. Then, for the $\mu_g(r,T_D)$ 
which corresponds to $\min \Delta_\mu$, we compute the theoretical 
frequency $F^{\rm cc}_{th}$, equate it to the observed one, $F^{\rm 
cc}_{obs}$, and derive $\beta$. After that we repeat the numerical 
solution of the above equations with the renewed $\beta$ again for a set 
of $\nu$, determine the new $\nu$ corresponding to $\min \Delta_\mu$, 
and calculate the corrected $\beta$. Such a procedure is repeated 
iteratively until convergence is reached. For the starting value of 
$m^{cc}=$~70 M$_{\odot}$ the free parameters inferred are as follows: 
$\nu~=~0.07421$ M$_{\odot}^{-0.5}$ pc$^{-2}$ Gyr$^{-1}$; 
$\beta~=~0.02309$ M$_{\odot}^{-0.5}$ pc (their corrected values will be 
given in \S3.1). In Fig. 4 we show the temporal evolution of the gaseous 
density for the final parameters after their corrections (see \S 3.1 for 
details).\footnote{In our previous papers (e.g., AMR and AMK) we 
followed {\it literatim}, the idea of Oort (1974) and introduced the 
factor $|\Omega - \Omega_P|$ into the enrichment rates $E^{\rm cc}_j$ 
and $E^{\rm Ia-P}_j$ as an additional multiplier retaining the same 
Schmidt-like representation for SFR ($\propto \mu_g^k$) for stars of all 
masses. But in the present paper, we explicitly split the star formation 
rate into low and high mass components and approximate them by different 
functional representations (cf. equations 5 and 6). As a consequence, we 
derive a radial density distribution for the gas which differs 
significantly from that which was obtained in our earlier work. In 
particular, we now predict a surface density distribution with a `hole' 
in the center which resembles closely the observed one of Dame (1993). 
Such quantitative and qualitative changes have entailed the changes in 
the sought-for parameters.}

\begin{figure}
\includegraphics {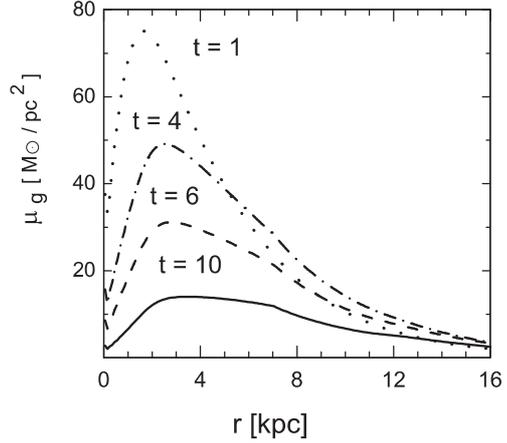}
\caption{Temporal evolution of the radial gas surface density 
distribution $\mu_g(r,t)$; see text for details.} 
\label{f4}
\end{figure}

\subsection {{\it Rates of SNeIa Events}}

`Tardy' SNeIa are not concentrated in spiral arms, and we describe their 
event rate usually as:
\begin{equation}
R^{\rm Ia-T}(r,t)= \zeta
\int\limits_{\tau_S}^{t}{\psi_{L}(r,\,t-\tau)D_T(\tau)d\tau},
\end{equation}
\noindent
where $D_T$ is the part of the `{\it Delay Time Distribution}' (DTD) 
function for the tardy sub-population (Mannucci et~al. 2006; Matteucci 
et~al. 2006; Maoz et~al. 2010) and $\tau$ is the time delay between the 
birth of a corresponding progenitor and its explosion (below we give the 
results for the smooth DTD function of Maoz et~al. 2010; the results for 
the bimodal one of Mannucci et~al. 2006 are close to that described 
below). The constant $\zeta$ is derived by means of computing $F^{\rm 
Ia-T}_{th}$ and equating it to $F^{\rm Ia-T}_{obs}$ .

Since the prompt sub-population is concentrated in spiral arms, their 
event rate is functionally comparable to that of the CC~SNe:
\begin{equation}
R^{\rm Ia-P}= \gamma \psi_{H}(r,t)
\int\limits_{\tau_8}^{\tau_S}{D_P(\tau)d\tau},
\end{equation}
\noindent
where $D_P$ is the part of DTD function corresponding to SNeIa-P (see 
Maoz et~al. 2010 and AMK) and $\gamma$ is a normalizing constant which is 
determined by means of equating the theoretical and observed frequencies 
for SNeIa-P events.

According to Li et~al. (2011) the observed frequency of SNeIa today, 
$F^{\rm Ia}_{obs}$, is $\sim$0.54 per century. Using their ratio: 
$$F^{\rm Ia-P}_{obs} / F^{\rm cc}_{obs} \approx~0.187 $$ we find: 
$F^{\rm Ia-P}_{obs}\approx~0.43$ and $F^{\rm Ia-T}_{obs}\approx~0.11$ 
per century. The above authors note that their values may have 
systematic uncertainties up to a factor $\sim$2. As such, the errors in 
observations carry the most weight in the search for the optimal 
parameters.

\subsection {\it Statistical Method for Deriving the Radial Distribution 
of Oxygen and Iron}

Having the dependence $\mu_g(r,t)$ for fixed parameters $P^{\rm i}_j$, 
we may solve numerically the chemical evolution equations (2). To derive the 
target parameters we minimize the discrepancy function $\Delta_X$ over 
$P^{\rm i}_j$:
\begin{equation}
\Delta^2_X = \frac{1}{n-p}\sum_{k=1}^n\{(\langle[X/H]^{obs}\rangle_k - 
[X/H]^{th}_k)w_k\}^2,
\end{equation}
\noindent
where the superscript `obs' corresponds to the observational data and 
`th' to the theoretical data, $w_k$ is a weight (we assume it to be 
inversely proportional to the length of the error-like bar in the $k$th 
bin, as in Fig.~1), $n=20$ is the number of bins, $p$ is the number of 
free parameters, and the summation is taken over all $k$th points within 
the adopted Galactocentric radius range.

Let us consider the procedure now in somewhat more detail. Since oxygen 
is mainly produced by CC~SNe and we can neglect the contributions from 
other sources (see AMR and AMK, and references therein), the 
calculations for this element are performed separately.  To find the 
minimum of the discrepancy function, we solve numerically equation (2) 
for a set of $P^{\rm cc}_O$ (and $E^{\rm Ia}_O = 0$), varying its value 
over a wide range at some step, and then compute the theoretical 
distribution at $t = T_D$, for each ejected mass of oxygen. Further, by 
means of equation (12), we derive the dependence of $\Delta_O$ as a 
function of $P^{\rm cc}_O$ and search for its minimum. Hence, by such 
means, we derive the first free parameter, $P^{\rm cc}_O$, independent 
of the others (in this case, $p=1$).

In contrast, iron is produced by three sources -- CC~SNe, SNeIa-P, and 
SNeIa-T. Tardy SNeIa do not concentrate within spiral arms; rather, they 
contribute to the radial distribution of iron with an approximately 
constant gradient. Unlike tardy SNeIa, CC~SNe and prompt SNeIa are 
concentrated in spiral arms and are responsible for the inflection in 
the radial abundance gradients. It is obvious from equations (9) and 
(11) that the rates $R^{\rm cc}$ and $R^{\rm Ia-P}$ have very close 
functional dependencies upon $r$. Therefore, it is impossible to 
distinguish their contributions by means of statistical analyses, if we 
do not introduce any {\it a priori} assumptions (in statistics, this 
problem is referred to as the {\it multi-collinearity} problem). 
Therefore, we can only indicate the upper values of $P^{\rm cc}_{Fe}$ 
and $P^{\rm Ia-P}_{Fe}$.

To describe the algorithm for the search of the mean ejected masses of 
iron we write the corresponding enrichment rate of the Galactic disk by 
the element, $E_{Fe}$, in explicit form:
\begin{equation}
E_{Fe}  = P_{Fe}^{\rm cc}R^{\rm cc} + P_{Fe}^{\rm Ia-P}R^{\rm Ia-P} + 
P_{Fe}^{\rm Ia-T}R^{\rm Ia-T}.       
\end{equation}
\noindent

Next, suppose that, say, $P_{Fe}^{\rm cc}=0$; substituting $E_{Fe}$ into 
equation (2), we solve for a set of pairs $(P_{Fe}^{\rm Ia-P},\, 
P_{Fe}^{\rm Ia-T})$, compute the radial distribution of iron at $t=T_D$, 
and find the minimum of the discrepancy $\Delta_{Fe}$ which gives the 
best parameters of $P_{Fe}^{\rm Ia-P}$ and $P_{Fe}^{\rm Ia-T}$.

At the second step, we assume that $P_{Fe}^{\rm Ia-P}=0$, again solve 
numerically equation (2) for a set of pairs $(P_{Fe}^{\rm cc},\, 
P_{Fe}^{\rm Ia-T})$, compute the radial distribution of iron, and seek 
the minimum of the discrepancy $\Delta_{Fe}$ which now gives the best parameters of $P_{Fe}^{\rm cc}$ and $P_{Fe}^{\rm Ia-T}$. Thus, we derive the best value for $P_{Fe}^{\rm Ia-T}$ and for $P_{Fe}^{\rm cc}$ and $P_{Fe}^{\rm Ia-P}$ (at this step in the treatment of iron, $p$=2).

The above estimates for the mean masses of iron ejected by CC~SNe or 
SNeIa-P are only upper limits since it is impossible to separate their 
simultaneous contributions, unless we possess additional information.

\section {Results and Discussion}
\subsection {{\it Oxygen}}
 
As noted in \S2.2, we performed several experiments with various initial 
values for $Z_O(t=0)$, the results for which show this choice does not 
impact on our conclusions. The application of the aforementioned modelling 
of the observed oxygen distribution results in a predicted mean mass of 
$P_O^{\rm cc}\approx~0.28\pm 0.01$~M$_{\odot}$ ejected per CC~SNe 
event.\footnote{The associated random error is much less than the 
error due to the uncertainty in the observed frequencies of SNe events (Li 
et~al. 2011). Hence, we neglect the random error in $P_O^{\rm cc}$.} This 
value differs significantly from the ones (1.8~--~3.7~M$_{\odot}$) 
proposed by Woosley \& Weaver (1995, hereafter WW95), Tsujimoto et~al. 
(1995, hereafter T95), and Thielemann et~al. (1996).\footnote{The low 
inferred value for $P_O^{\rm cc}$ is, though, more comparable to the 
yields of Arnett (1991) and Langer \& Henkel (1995); see Table~1 of 
Gibson, Loewenstein \& Mushotzky (1997).}  We now discuss the reason for 
this divergence and the consequence of the predicted lower mean ejected 
mass of oxygen per CC~SNe event.

To estimate $P_O^{\rm cc}$, one needs the mass-dependent oxygen yields 
from the relevant stellar evolution grid, and then to convolve that with 
an appropriate IMF and its associated upper limit for stars ending their 
lives as CC~SNe (e.g., $m_U \sim$~50 - 70 M$_\odot$, as in T95). In \S2.3, 
we also assumed that in equation (9), $m^{cc}_U \equiv m_U$, but this 
adopted solely as an initial `guess'.  We now define $m^{cc}_U$ more 
accurately. For this, we use the procedure close to the one described by 
Gibson (1998). There, the mean mass of oxygen, $<M_O>$, ejected per CC SNe 
event (whose progenitor had the initial mass $m$), was represented as:

\begin{equation} 
<M_O> = \frac{\int\limits_{10}^{m^{\rm cc}} M_O(m)\phi(m) dm}
       {\int\limits_{m_L^{\rm cc}}^{m^{\rm cc}} \phi(m) dm}.
\end{equation}

\noindent
Let us discuss the lower limits in the above integrals. As argued by 
WW95 (see also T95), CC~SNe of masses approximately in the interval 
8--10~M$_{\odot}$ contribute very little to the production of elements 
like oxygen. That is why the lower limit in the integral in the numerator 
is set to 10~M$_{\odot}$. T95 use the same value (10~M$_{\odot}$) for the 
lower limit in the integral entering the denominator in equation (14). We 
believe though that $m^{\rm cc}_L$ must be equal to the lower limit in the 
integral representing the rate of CC~SNe events (see equation 9), since 
the stars in the mass range 8--10~M$_{\odot}$ contribute to the observed 
frequency of the corresponding CC~SNe events (Li et~al. 2011; Smartt 
et~al. 2009). In other words, these stars influence the mean ejected mass 
of oxygen through the rate of CC~SNe events. Hence, we should impose 
$m^{\rm cc}_L = 8$ M$_{\odot}$. 

Using for $M_O(m)$ the yields published in literature, we compute the 
dependence of $<M_O>$ as a function of $m^{\rm cc}$ and, by equating $<M_O> = 
P^{\rm cc}_O$, we find the new upper initial mass, $m^{\rm cc}_U$, of a 
star which can explode as a CC~SNe. For $P_O^{\rm cc}=~0.28$~M$_{\odot}$ 
and the yields of T95, the `corrected' initial mass comes to $m^{\rm 
cc}_U \approx 23.5$~M$_{\odot}$ (see Fig. 5).

\begin{figure}
\includegraphics {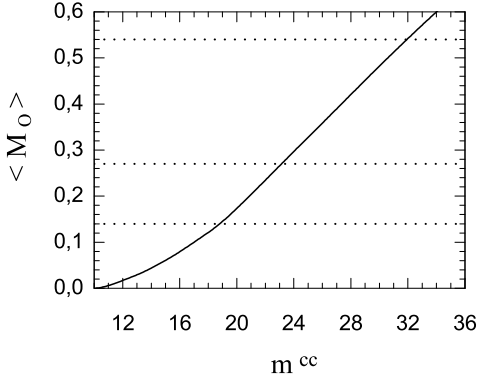}
\caption{The dependence of $<M_O>$ on $m^{\rm cc}$ for the yields of T95.}
\label{f5}
\end{figure}

Further, since the new $m^{\rm cc}_U$ differs from the initial one 
(70~M$_{\odot}$), we have to launch the next iterative step. For this, we 
substitute the new value $m^{\rm cc}_U=$~23.5~M$_{\odot}$ into equation 
(9) and repeat the procedure from the beginning, to solve anew the 
equations for evolution of $\mu_g$, re-determine the constants $\nu$, 
$\beta$, and $P_O^{\rm cc}$ and derive the improved value for $m^{\rm 
cc}_U$. Our calculations show that the corrected values of the constants 
are: $\nu~=$~0.07177~M$_{\odot}^{-0.5}$ pc Gyr$^{-1}$, $\beta 
=$~0.03032~M$_{\odot}^{-0.5}$ pc (the dependence of $\mu_g(r,t)$ for these 
final parameters is shown in Fig.~4), and $P_O^{\rm cc} 
\approx$~0.27~M$_{\odot}$. The corresponding mean ejected mass of oxygen 
happens to be slightly less than the previous value: $m^{\rm 
cc}_U=$~23.1~M$_{\odot}$ (see Fig. 5). To estimate the scatter in $m^{\rm 
cc}_U$ due to the uncertainty in the observed frequency of CC SNe events 
we adopt a value for $F^{\rm cc}_{ob}$ two times greater (or lower) than 
the `best' value of Li et~al. (2011). Correspondingly, $P^{\rm cc}_O$ will 
be a factor of two lower (or greater) than the above derived value. By 
means of Fig. 5, we find: $m^{\rm cc}_U=23.1^{\rm +8.9}_{\rm 
-4.3}$~M$_{\odot}$. If we use the yields of Hirschi et~al. (2005), $m^{\rm 
cc}_U$ will be systematically lower by $\sim$2~M$_\odot$.\footnote{Note that the changes in the upper initial mass for the exploding CC~SNe 
follow from the non-linear dependence of the mean ejected mass of oxygen on 
$m^{\rm cc}_U$.} The theoretical radial distribution of oxygen in the 
Galactic disk, superimposed on the observed one, is shown in Fig. 1.

The above upper initial mass for CC SNe progenitors is close to that 
favoured by Maeder (1992) and Heger et~al. (2003) who proposed an upper 
initial mass limit for exploding SNe to be $\sim$20--25~M$_{\odot}$. 
Moreover, on the basis of observations, Smartt et~al. (2009) insist that 
the progenitors of exploded CC~SNe have masses below $\sim$17~M$_{\odot}$. 
Eldridge et~al. (2013) also do not find evidence for the existence of very 
massive stars ($\sim$100 M$_{\odot}$) which might be considered as 
progenitors for observed SNe Ib/c. Finally, Kochanek et~al. (2008) suggest 
that stars more massive than the above transform directly to black holes 
at the end of nuclear burning without exploding (or undergo a backward 
explosion).

For our theory it is crucial that, independent of the final stages of 
stellar evolution, stars with very large initial masses do not 
necessarily explode as CC~SNe. In our picture, if they did, they would 
supply too much oxygen to the Galactic disk. In our approach, we are not 
faced with the problem of an excessive frequency of CC~SNe events, noted 
by Brown \& Woosley (2012), since the observed frequencies are built 
into the theory at the time of disk formation.

\subsection {{\it Iron}} 

After improvements associated with a more precise definition of $m^{\rm 
cc}_U$, we derive the final values $\gamma =$~0.0061 and $\zeta =$~0.0007. 
The mean mass of iron ejected per SNeIa~-~T event is $P_{Fe}^{\rm Ia-T} 
\approx 0.58 \pm 0.20$ M$_{\odot}$ and it is approximately the same, 
independent of the type of short-lived SNe excluded (core-collapse or 
prompt SNeIa, see \S2.5). This is expected, since tardy SNeIa are 
responsible for the quasi-linear radial distribution of iron along the 
Galactic disk.

Further, if $P_{Fe}^{\rm Ia-P}=0$, then the mean mass of iron ejected per 
CC~SNe event is $P_{Fe}^{\rm cc} \approx 0.04 \pm 0.01$ M$_{\odot}$ 
whereas for the case $P_{Fe}^{\rm cc}=0$, the corresponding mass ejected 
per prompt SNeIa event is $P_{Fe}^{\rm Ia-P} \approx 0.23 \pm 0.06$ 
M$_{\odot}$ (for the corresponding theoretical radial distribution of 
iron, see Fig. 1).\footnote{The temporal evolution of radial abundance 
gradients is beyond the scope of our analysis, but a deeper investigation 
into the issues can be found in Pilkington et~al. (2012) and Gibson et~al. 
(2013).}

As mentioned in \S1, SNeIa are usually associated with long-lived objects 
and the yields of iron, estimated on the basis of a theory of pre-SNeIa 
evolution (e.g., Nomoto et al. 1997; Iwamoto et al. 1999; Umeda \& Nomoto 
2002; Limongi \& Chieffi 2003; Kobayashi et al. 2006; Tominaga et al. 
2006; Woosley et al. 2007), are derived without separation of SNeIa into 
prompt and tardy sub-populations. Our estimates of the mean mass, ejected 
by prompt and tardy SNeIa per respective event, are the first indication 
that these sub-populations do produce different amounts of iron. If the 
events arise in similar conditions (white dwarfs -- the progenitors of 
SNeIa -- have masses in a very narrow range), how can this be understood? 
Perhaps the discrepancy may be explained by the weak dependence between 
the peak luminosity and explosion kinetic energy for most SNeIa (Blondin 
et al. 2012), or perhaps, SNeIa-P undergo asymmetric explosions (Maeda et 
al. 2010)?

The low mean mass of iron, ejected per core-collapse SNe event 
($\sim$0.04$\pm$0.01~M$_{\odot}$), is close to the observed one given by 
Smartt et al. (2009) - 0.01 -- 0.03~M$_\odot$ - for carefully measured 
CC~SNe. Hence, an even more challenging assumption might be to speculate 
that \it perhaps \rm the nature of prompt SNeIa is not associated with 
white dwarfs, but rather with CC~SNe. Indeed, as it was noted in \S1, 
observations show that `young' SNeIa-P are brighter than `old' tardy 
SNeIa, but according to our results prompt SNeIa produce (per event) about 
2.5 times less iron than tardy SNeIa. How can this (`inverse correlation') 
be explained if the main source of SNeIa luminosity is thought to be the 
decay of nickel to iron which is synthesized in a process with similar 
characteristics to that of the Chandrasekhar limit for white dwarfs? The 
low values of iron produced by CC~SNe and SNeIa-P make it tempting to 
ascribe credence to our suggestion. Besides, it is well known that, like 
core-collapse SNe, prompt SNeIa are concentrated in sites of star 
formation within spiral arms (Bartunov et al. 1994; Mannucci et al. 2005; 
Panagia et al. 2007; Li et al. 2011). Hence, we may assume that their 
progenitors were sufficiently massive stars. Moreover, in a recent paper 
Foley et~al. (2012) revealed (for the first time) strong outflows from 
SNeIa progenitors in late-type galaxies, but in early-type galaxies such 
outflows are extremely rare. If we recall that in early galaxies we 
observe `old' (i.e., tardy) SNeIa but in late-type galaxies one sees 
`young' (prompt) SNeIa, we might conclude that the phenomenon of Foley et 
al. provides a `link' between SNeIa-P and CC~SNe. Further, it is worth 
noting the work of van~Rossum (2012) who, on the basis of recent data, 
demonstrates that some features in SNeIa spectra may be interpreted 
incorrectly (e.g., emission may be interpreted as absorption). In some 
cases, this may mislead the classification of SNe type.

Our experiments enable us to estimate the total number of SNe, $N^{\rm 
i}$, exploded in the Galaxy throughout its life: $N^{\rm 
cc}$$\sim$12.2$\times$10$^8$; $N^{\rm Ia-P}$$\sim$2.3$\times$10$^8$; 
$N^{\rm Ia-T}$$\sim$0.2$\times$10$^8$. From these, we can estimate that 
short-lived SNe (CC or Ia-P) supply to the Galaxy $\sim$~85\% of its iron. 
Correspondingly, SNeIa-T supply $\sim$~15\% of the iron. 

The low value inferred for the amount of iron supplied to the Galaxy by 
the long-lived sub-population of SNeIa differs from the one typically 
discussed in the literature. Indeed, prior to the most recent 
decade, SNeIa were considered to be (essentially) exclusively long-lived 
objects, and under such a picture, they were though to supply $\sim$60\% 
of the iron to the Galaxy (Gibson 1998; Matteucci 2004). This 
stance has changed since the discovery of the two sub-populations of 
SNeIa (short- and long-lived).\footnote{According to our definition 
(\S2.3), SNeIa of ages more than $\sim$400~Myr (Matteucci \& Greggio 
1986) are considered to be SNeIa-T.} As was shown by AMK, the 
relative portion of iron supplied to the Galaxy by tardy SNeIa reduced 
to $\sim$~35\%, since the rapid SNeIa (and/or CC~SNe) also contribute to 
iron synthesis. According to our workd here, the contribution of 
long-lived SNeIa to the disk's iron enrichment appears a factor of 
$\sim$2 lower than even this reduced fraction. This is mainly a 
consequence of our new model for the SFR function, $\psi$, which we 
explicitly split into low and high massive star formation rates, the 
mathematical representations for these two parts being very different. 
This modification drives the relatively low amount of iron both 
contributed to the Galactic disk and ejected per tardy SNeIa event 
$\sim$0.4 - 0.8~M$_{\odot}$.

Despite the fact that our yields are derived in the context of a 
galactic chemical evolution model for the Galactic disk, it is tempting 
to extend their application to the stellar halo. Indeed, such stars are 
to be enriched primarily by earlier generations of short-lived SNe.  Can 
we explain the enhanced ratios of [$\alpha$/Fe] in halo stars via the 
use of our empirically inferred yields? If we assume that the enrichment 
seen in the oxygen and iron abundances of halo stars was due to CC~SNe 
only, one can estimate the expected ratio as being $[O/Fe]=log(P^{\rm 
cc}_O/m_O)- log(P^{\rm cc}_{Fe}/m_{Fe})-log(O/Fe)_{\odot}$, where $m_O$ 
and $m_{Fe}$ are the atomic masses of oxygen and iron, respectively, and 
for the solar ratio in our work we adopt $log(O/Fe)_\odot$=1.19 (in 
all our work, we employ the solar scale of Asplund et~al. 2009, since 
this scale was also used in the determination of the elemental patterns 
from the aforementioned Cepheid samples). For the above yields, we 
derive [O/Fe]=$+$0.18 (for the lower value of $P^{\rm cc}_{Fe} = 0.03$ 
M$_{\odot}$, the ratio comes to [O/Fe]=$+$0.31). While our framework was 
not created with the stellar halo in mind, it is at least re-assuring to 
note that under the assumption of rapid enrichment, our inferred 
estimates of the resulting [$\alpha$/Fe] are not dissimilar to the 
$\sim$$+$0.2$-$0.3~dex $\alpha$-enhancements seen in typical metal-poor 
environments.  Further, we should re-iterate that our quoted mean mass 
of iron ejected per CC~SNe event is technically only an upper limit. We 
hope that a more careful analysis which explicitly takes into account 
the evolutionary history of halo, like in Renda et~al. (2005), will 
enable us to better apply our preliminary work outside of the regime for 
which it has been optimised here.  Stronger observational constraints on 
various sub-types of SNe, particularly in distant galaxies, are also 
critical, as (for example) the data of Li et~al. (2011) provides 
insights really only into the SNe rates today, rather than as a function 
of time.

There may in fact be another way to explain the enhanced $\alpha$/Fe 
ratio in halo stars: specifically, due to the very low metallicity 
($Z\sim0.001Z_{\odot}$) at the epoch of halo formation, the initial masses 
of CC~SNe may be higher (say, $\sim$40~M$_{\odot}$) than the masses of 
CC~SNe exploding in the Galactic thin disk. According to Maeder (1992) 
such CC~SNe progenitors produce several times more oxygen than their 
counter-parts at solar metallicity.

\section{Conclusions}

Our goal was to derive the mean masses of oxygen and iron ejected per each 
sub-type of SNe, and, as a consequence, the constraints on some properties 
of their progenitors. For this, we refined the statistical analysis of the 
observed (non-linear) radial distributions of elements in the Galactic 
disk previously developed by AMR and AMK.  Breaking with the traditional 
approach of adopting tabular nucleosynthetic yields in the chemical 
evolution modeling, we instead derive them \it a posteriori\rm, within our 
framework (\S2).

The results are as follows:
\begin{itemize}
\item The mean mass of oxygen ejected per CC~SNe event is $\sim$0.27 
M$_{\odot}$, while the upper value for the mass of iron, ejected by this 
type of SNe is $\sim$0.04 M$_{\odot}$.
\item The mean mass of iron ejected by tardy SNeIa (progenitors which 
are long-lived with ages $\simgt$100 Myr, up to several Gyrs, and which 
are not concentrated in spiral arms) is $\sim$0.58 M$_{\odot}$ per 
event. Conversely, prompt SNeIa (progenitors which are short-lived with 
ages $\simlt$100~Myr and which are concentrated within spiral arms) 
supply to the Galactic disk $\leq$0.23 M$_{\odot}$ per event (on 
average).
\item Short-lived SNe (core-collapse and prompt Ia) supply to the 
disk $\sim$85\% of the Galaxy's iron.
\end{itemize}

The low amount of oxygen ejected per CC~SNe leads us to the conclusion 
that the maximum initial mass of exploding CC~SNe progenitors is 
$\leq$~23$^{\rm +8.9}_{-4.3}$ M$_{\odot}$. This result supports the 
inferences of Heger et al. (2003), Kochanek et al. (2008), and Smartt et 
al. (2009) -- i.e., that stars of initial masses greater than about 17~-- 
25 M$_{\odot}$ do not explode and, as such, they do not return the bulk of 
their newly synthesized material to the ISM. Perhaps, they collapse to 
black holes shortly after their birth without explosion or undergo a 
backward explosion. As a consequence, they do not take part in Galactic 
nucleosynthesis (except perhaps through their pre-SN stellar wind 
mass-loss).

The mean mass of iron ejected by the typical CC~SNe ($\leq$~0.04 
M$_{\odot}$) is close to the empirical values favoured by Smartt et al. 
(2009). The mean masses of iron ejected by prompt and tardy SNeIa appear 
to be different. It is surprising that, being older and less energetic, 
the tardy ones produce about 2.5 times more iron (per event) than prompt 
SNeIa, which are younger and more energetic objects (Gonzalez-Gaitan 
et~al. 2011). This suggests that, perhaps, in their nature, SNeIa-P are 
closer to core-collapse SNe than to those whose explosions are associated 
with the thermonuclear burning of a Chandrasekhar mass white dwarf.

\section*{Acknowledgments}
The authors wish to thank A.~Zasov, S.~Blinnikov, D.~Tsvetkov, and the 
anonymous referee, for their valuable guidance. This work was supported by 
the Ministry of Education \& Science of the Russian Federation 
(14.A18.21.0787 \& 14.A18.21.1304) and the grant of the Southern Federal 
university. BKG acknowledges the support of the UK Science \& Technology 
Facilities Council (ST/J001341/1). IAA thanks the Russian Funds for Basic 
Research scheme (12-02-90701-mob-st).

\label{lastpage}
\end{document}